\documentstyle[11pt,aaspp4]{article} 
\begin{document}
\title{Viscosity in X-ray clusters: Braginskii over 5} 
\author{Andrei Gruzinov}
\affil{Physics Department, New York University, 4 Washington Place, New York, NY 10003}

\begin{abstract}

We argue that it is currently impossible to simulate X-ray clusters using correct equations, because even the MHD description is not applicable. But since fluid simulations actually reproduce observations quite well, one may try to improve the fluid codes by including molecular transport of heat and momentum. We calculate the effective molecular viscosity for the simplest model of magnetic field and obtain 1/5 of the Braginskii value, similar to 1/3 of Spitzer for the heat conduction. This is large enough to noticeably damp the X-ray cluster turbulence.

\end{abstract}

\section{Introduction}

To turn X-ray clusters into a precise probe of cosmology, one needs to simulate the clusters numerically. This is being done (eg. Nagai, Kravtsov, Vikhlinin 2006, Sijacki, Springel 2006 and references therein) but there are some fundamental problems with these or any potentially doable simulations (\S 2). 

Given that the first principle numerical simulation is not possible, one can still try to use some effective fluid description. And in fact remarkable agreement with observations was already achieved (Nagai, Kravtsov, Vikhlinin 2006). Further improvement, which is needed for cosmology, may come when the missing physics is included. One obvious thing is to include molecular heat conduction and viscosity (eg. Gruzinov 2002, Sijacki, Springel 2006). 

It has been shown that in the simplest model with random magnetic fields, the effective molecular heat conduction equals 1/3 of the unmagnetized (Spitzer) value. Here we show (\S3) that the same magnetic field model gives effective molecular viscosity equal to 1/5 of the unmagnetized (Braginskii) value. Of course, practical numerical simulations should be done  using some fudge factors correcting the Braginskii-Spitzer transport coefficients (Sijacki, Springel 2006). We are merely pointing out that 1/3 and 1/5 are, in principle, realizable values of the fudge factors.

\section{Fluid description of X-ray clusters}

Fluid description is not applicable to the cluster plasma. Here by the fluid description we mean ideal or non-ideal, pure hydro or MHD, or even Braginskii (1965) non-ideal multiple-fluid MHD. Collisional Vlasov (Maxwell electrodynamics plus Boltzmann equation with collisions) is, of course, valid, but it requires particle simulations which seem to be well out of reach.

The problem with the fluid description is the magnetic field. Even if it turns out that the magnetic field is very weak (and this is a big if given the desired $\lesssim 1$\% accuracy), one must still calculate it, because the magnetic field determines the plasma transport coefficients. And it is known that molecular heat conduction and viscosity, if close to the Braginskii-Spitzer values, have a significant effect on the cluster (eg. Gruzinov 2002, Sijacki, Springel 2006). 

If magnetic fields must be calculated, why cannot one calculate those using the Braginskii equations? This is best explained by the frozen-in law. 

According to Maxwell equations (Faraday law), $\partial _t{\bf B}=\nabla \times {\bf E}$. For the cluster plasma, the electric field component parallel to the magnetic field must nearly vanish (plasma electric conductivity along the magnetic field is very high). But if ${\bf E}$ is perpendicular to ${\bf B}$, one can introduce an auxiliary vector field ${\bf v}$, such that ${\bf E}={\bf v}\times {\bf B}$, and then write the Faraday law as 
\begin{equation}\label{frozen-in}
\partial _t{\bf B}=\nabla \times ({\bf v}\times {\bf B}).
\end{equation}
This equation means that the magnetic field ${\bf B}$ is frozen into the auxiliary field ${\bf v}$. Due to the frozen-in law, the magnetic field topology (in particular the magnetic helicity $\int d^3r{\bf A}\cdot {\bf B}$, where ${\bf A}$ is the vector potential, ${\bf B}\equiv \nabla \times {\bf A}$) is conserved. 

If one could calculate the auxiliary field ${\bf v}$ (only the components of {\bf v} perpendicular to ${\bf B}$ are needed and defined), the frozen-in equation (\ref{frozen-in}) would give the magnetic field. On large scales, larger then the plasma mean free path 
\begin{equation}
\lambda \sim ~\left( {T\over 3 {\rm keV}}\right) ^2~\left( {10^{-3} {\rm cm}^{-3}\over n}\right) ~{\rm kpc}~, 
\end{equation}
the auxiliary field ${\bf v}$ coincides with the plasma velocity. The problem is that the magnetic field still exists and evolves iso-topologically on length scales of order $\lambda$ and smaller. For example, on the length scale equal to the plasma mean free path, the magnetic field damping time $t_{md}$ (due to finite electric conductivity) can be written as 
\begin{equation}
{t_{md}\over \tau _e}\sim {1\over nr_e^3}\left( {T\over m_ec^2}\right) ^4\sim 10^{32},
\end{equation}
where $\tau _e$ is the electron collision time, and $r_e$ is the classical electron radius. This number is too big to be taken literally. Kinetic plasma instabilities will probably develop and change the field --  but the problem is that we don't know exactly how the field will evolve on these length scales.

We can also state the problem with the fluid description in a slightly different language. As a warm-up, consider fluid description of an unmagnetized gas. The gas is characterized by the mean free path $\lambda$ and the sound speed $c_s$. Suppose a flow of velocity $\sim V<c_s$ is imposed at length scales $\sim L\gg \lambda$. According to the fluid description, Kolmogorov cascade to smaller scales develops, with characteristic velocity $v_l\sim V(l/L)^{1/3}$ on all scales $l$ from the inertial range $l_d<l<L$. Here the dissipation scale $l_d$ is given by the equality of the eddy frequency and the damping rate: $\nu /l_d^2\sim  v_{l_d}/l_d$, where $\nu \sim c_c\lambda$ is the molecular viscosity. It follows that $l_d/\lambda \sim (c_s/V)^{3/4}(L/\lambda)^{1/4}~\gg ~1$. This means that the fluid description is consistent -- the flow is fully damped on scales much larger than the mean free path $\lambda$. 

Now consider magnetic fields in the cluster plasma, assuming Braginskii MHD description. We don't know how to calculate magnetic field cascade to small scales. We will simply underestimate the characteristic ``eddy'' frequency replacing it with the large scale frequency $V/L$. Then the magnetic dissipation scale $l_{md}$ is given by $\nu _{m}/l_{md}^2\sim V/L$, where $\nu _{m}\sim c^2/\sigma$ is the magnetic diffusivity, and $\sigma \sim c^2r_en\tau _e$ is the electric conductivity. This gives 
\begin{equation}
{l_{md}\over \lambda }\lesssim \left( {m_p\over m_e}\right) ^{1/4} \left( {V\over c_s}\right) ^{-1/2}\left( {L\over \lambda }\right) ^{1/2}\left( {T\over m_ec^2}\right) ^{-2}(nr_e^3)^{1/2} ~\sim 10^{-14}.
\end{equation}
Again this is too small -- kinetic effects will turn on and cut off the cascade. But we don't know what actually happens.

The problem of the cluster magnetic fields and molecular transport is related to the well-know problem of high-energy astrophysics -- effective magnetohydrodynamical description of collisionless plasma. It might turn out that MHD (with anisotropic molecular transport) can mimic the outcome of kinetic instabilities, but this has not been demonstrated. 

\section{Effective viscosity in a random time-varying magnetic field}

We start by repeating our (trivial) argument for the Spitzer over 3 heat conduction. When the same procedure is applied to viscosity, one gets Braginskii over 5. 

{\bf Heat conduction:} Heat conduction along the magnetic field is equal to $\kappa _S \nabla _\parallel T$, were $\nabla _\parallel T$ is the temperature gradient along the field, $\kappa _S$ is the Spitzer thermal conductivity (Braginskii 1965). If the magnetic field evolves randomly and fast \footnote{Say, the cluster plasma is being crisscrossed by about a hundred galaxies. The galaxies which have retained their plasma entrain the magnetic field, forcing reconnection on the short -- galaxy crossing -- time scale.}, one should average over the magnetic field direction. Averaging the heat flux over the random magnetic field direction ${\bf b}$ gives Spitzer over 3: 
\begin{equation}
<{\bf q}>=-<{\bf b}\kappa _S {\bf b}\cdot \nabla T>= -{\kappa _S\over 3} \nabla T.
\end{equation}

{\bf Viscosity:} Now we should average the momentum flux tensor
\begin{equation}\label{pi}
\pi _{ij}=-\eta _B {3\over 2}(b_ib_j-{1\over 3}\delta _{ij})b_nb_mW_{nm},
\end{equation}
where the rate of strain tensor is 
\begin{equation}
W_{nm}\equiv \partial _nV_m+\partial _nV_m-{2\over 3}\nabla \cdot {\bf V}\delta _{nm},
\end{equation}
and ${\bf V}$ is the plasma (ion) velocity. Averaging eq.(\ref{pi}) over the random unit vector ${\bf b}$ gives Braginskii over 5:
\begin{equation}
<\pi _{ij}>=-{\eta _B\over 5}W_{ij}
\end{equation}

We now re-derive the suppression factors 1/3 and 1/5 using the heat production rate. For heat conduction, $Q\propto (\nabla T)^2$ without magnetic field, and $Q\propto ({\bf b}\cdot \nabla T)^2$ with magnetic field. Averaging over ${\bf b}$ gives the factor 3 suppression. 

For viscosity, $Q=(\eta _B/2) ({\bf W})^2$, without magnetic field. With non-zero magnetic field, the heat production rate must be proportional to $({\bf b}\cdot {\bf W}\cdot {\bf b})^2$, because all other components of the rate of strain can be represented by the velocity field with a gradient perpendicular to ${\bf b}$, giving a suppressed transport and damping. Then $Q= (\eta _B/2)(3/2)({\bf b}\cdot {\bf W}\cdot {\bf b})^2$. Here the normalization coefficient $3/2$ can be obtained by taking the rate of strain tensor ${\bf W}={\rm diag}(-1/2,-1/2,1)$ for ${\bf b}=(0,0,1)$, and noting that for this geometry there should be no suppression of heating (without collisions, parallel and perpendicular temperatures change in the same way with and without magnetic field). Averaging the magnetized heating rate over ${\bf b}$ gives the factor 5 suppression.

\acknowledgements

I thank Alexey Vikhlinin for discussions. This work was supported by the David and Lucile Packard foundation.


\begin{references}

\reference{} Nagai, Kravtsov, Vikhlinin 2006, astro-ph/0611013
\reference{} Sijacki, Springel 2006, astro-ph/0610907
\reference{} Gruzinov 2002, astro-ph/0203031

\end{references}
\end{document}